\documentclass[preprint,superscriptaddress]{revtex4}
\usepackage{graphicx}
\usepackage{array}
\usepackage{amssymb}
\usepackage{amsfonts}
\usepackage{amsmath}
\usepackage{mathrsfs}
\usepackage{color}
\usepackage{booktabs}
\usepackage{threeparttable}
\usepackage{multirow}
\usepackage{subfigure}
\usepackage{times}
\usepackage{epsfig}
\usepackage{threeparttable}
\usepackage{chngpage}
\usepackage{latexsym}

\begin{document}

\title{Reconstructing Propagation Networks with Natural Diversity and
Identifying Hidden Source}

\author{Zhesi Shen}
\affiliation{School of Systems Science, Beijing Normal University,
Beijing, 100875, P. R. China}

\author{Wen-Xu Wang}\email{wenxuwang@bnu.edu.cn}
\affiliation{School of Systems Science, Beijing Normal University,
Beijing, 100875, P. R. China}
\affiliation{School of Electrical, Computer and Energy Engineering, Arizona
State University, Tempe, Arizona 85287, USA}

\author{Ying Fan}
\affiliation{School of Systems Science, Beijing Normal University,
Beijing, 100875, P. R. China}

\author{Zengru Di}
\affiliation{School of Systems Science, Beijing Normal University,
Beijing, 100875, P. R. China}

\author{Ying-Cheng Lai}
\affiliation{School of Electrical, Computer and Energy Engineering, Arizona
State University, Tempe, Arizona 85287, USA}
\affiliation{Department of Physics, Arizona State University,
Tempe, Arizona 85287, USA.}

\begin{abstract}
Our ability to uncover complex network structure and dynamics from data
is fundamental to understanding and controlling collective
dynamics in complex systems. Despite recent progress in this area,
reconstructing networks with stochastic dynamical processes
from limited time series remains to be an outstanding problem. Here we develop
a framework based on compressed sensing to reconstruct complex networks on
which stochastic spreading dynamics take place. We apply the methodology
to a large number of model and real networks, finding that a full reconstruction of
inhomogeneous interactions can be achieved from small amounts of polarized
(binary) data, a virtue of compressed sensing. Further, we demonstrate that
a hidden source that triggers the spreading process but is externally
inaccessible can be ascertained and located with high confidence in the
absence of direct routes of propagation from it. Our approach thus establishes
a paradigm for tracing and controlling epidemic invasion and information
diffusion in complex networked systems.
\end{abstract}

\maketitle

One of the outstanding problems in interdisciplinary
science is nonlinear and complex systems identification, prediction, and
control. Given a complex dynamical system, the various types of dynamical
processes are of great interest. The ultimate goal in the
study of complex systems is to devise practically implementable strategies
to control the collective dynamics. A great challenge is that the network
structure and the nodal dynamics are often unknown but only limited
measured time series are available. To control the system
dynamics, it is imperative to be able to map out the system details from
data. Reconstructing complex network structure and dynamics from data,
the inverse problem, has thus become a central issue in contemporary network
science and engineering~\cite{NPperspective,Systems_Biology,Timme:2007,
BL:2007,CMN:2008,RWLL:2010,LP:2011,HKKN:2011}. There are broad applications
of the solutions of the network reconstruction problem, due to the ubiquity of
complex interacting patterns arising from many systems in a variety of
disciplines~\cite{AB:2002,Newman:2003,NLMCH:2006,Newman:book}.

An important class of collective dynamics is
epidemic spreading and information diffusion in the human society or on
computer networks~\cite{epid1,epid2,epid3,epid4,epid5,epid6,epid7,epid8}.
The past decades have witnessed severe epidemic outbreaks at the global
scale due to the mutation of virus, including SARS~\cite{SARS1,SARS2},
H5N1~\cite{H5N11,H5N12}, H1N1~\cite{H1N11,H1N12} and the recent invasion
of H7N9 in eastern China~\cite{H7N91,H7N92}. Our goal is to reconstruct
the networks hosting the spreading process and identify the source of
spreading using limited measurements. This is especially challenging
due to (1) difficulty in predicting and monitoring mutations of deadly
virus and (2) absence of epidemic threshold in heterogeneous
networks~\cite{Onset1,Onset2,Onset3,Onset4}. Another example is rumor propagation
in the online virtual communities, which can cause financial loss or even
social instabilities, such as the 2011 irrational and panicked acquisition
of salt in southeast Asian countries caused by the nuclear leak in Japan.
In this regard, identifying the propagation network for controlling the
dynamics is of great interest. Another significant challenge in reconstructing
a spreading network lies in the nature of the available time series: they
are polarized, despite stochastic spreading among nodes. Indeed, the link
pattern and the probability of infection are encrypted in the binary status
of individuals, infected or not, analogous to the collapse of wave function to
one associated with some discrete quantum state induced by observation in
quantum mechanics.

There have been recent efforts in addressing the inverse problem of some
special types of complex propagation networks~\cite{invers1,invers2}.
In particular, for diffusion process originated from a single source,
the routes of diffusion from the source constitute a tree-like structure.
If information about the early stage of the spreading dynamics is available,
it would be feasible to decode all branches that reveal the connections
from the source to its neighbors, and then to their neighbors, and so on.
Taking into account the time delays in the diffusion process enables a
straightforward inference of the source in a complex network through
enumerating all possible hierarchical trees~\cite{invers1,invers2}. However,
if no immediate information about the diffusion is available, the tree-structure
based inference method is inapplicable, and the problem of network reconstruction
and locating the source becomes extremely challenging, hindering control
of diffusion and delivery of immunization. The loss of
knowledge about the source is common in real situations.
For example, passengers on an international flight can carry a highly
contagious disease, making certain airports the immediate neighbors of the
hidden source, which would be difficult to trace. In another example,
the source could be migratory birds coming from other countries or
continents. A general data-driven approach, applicable in such scenarios,
is still lacking.

In this paper, we develop a general theoretical framework to reconstruct
complex propagation networks from time series based on the compressed
sensing theory (CST)~\cite{CS1,CS2,CS3,CS4,CS5,CS6}, a novel optimization
paradigm for sparse-signal reconstruction with broad applications in
signal and image processing. Due to the striking characteristics of
CST such as the extremely low data requirement and rigorous guarantee of
convergence to optimal solutions, our framework is highly efficient and
accurate. However, casting the inverse problem into the CST framework
is highly nontrivial. Although CST has been used to uncover the nodal
interaction patterns for coupled oscillator networks or evolutionary games
from time series~\cite{CS:wwx1,CS:wwx2,CS:wwx3}, the dynamics of epidemic
propagation is typically highly stochastic with, for example, binary
time series, rendering inapplicable the existing CST-based formulation.
Further, despite the alternative sparsity enforcing regularizers and convex
optimization used in Ref.~\cite{SER:ML} to infer networks, CST has not been
applied to reconstructing propagation networks, especially when the
available time series are binary.
The main accomplishment of this work is then the development of a scheme
to implement the highly nontrivial transformation associated with the
spreading dynamics in the paradigm of CST. Without loss of generality, we
employ two prototypical models of epidemic spreading: classic
susceptible-infected-susceptible (SIS) dynamics~\cite{epid1} and
contact processes (CP)~\cite{CP1,CP2}, on both model and real-world
(empirical) networks. Inhomogeneous infection and recovery rates as
representative characteristics of the natural diversity are incorporated
into the diffusion dynamics to better mimic the real-world situation.
We assume that only binary time series can be measured, which
characterize the status of any node,
infected or susceptible, at any time after the outbreak
of the epidemic. The source that triggers the spreading process
is assumed to be externally inaccessible (hidden). In fact, one may
not even realize its existence from available time series.
Our method enables, based on relatively small amounts of data, a full
reconstruction of the epidemic spreading network with nodal diversity and
successful identification of the immediate neighboring nodes of the hidden
source (thereby ascertaining its existence and uniquely specifying its
connections to nodes in the network). The framework is validated with respect to
different amounts of data generated from various combinations of the
network structures and dynamical processes. High accuracy, high efficiency
and applicability in a strongly stochastic environment with measurement noise
and missing information are the most striking characteristics of our framework.
Thus broad applications can be expected in addressing significant
problems such as targeted control of disease and rumor spreading. \\

\noindent
{\bf \large{Results}} \\
\noindent
{\bf Compressed sensing.}
The general problem that CST addresses is to reconstruct a vector
$\mathbf{X} \in R^N$ from linear measurements $\mathbf{Y}$ about
$\mathbf{X}$ in the form
\begin{equation} \label{eq:cs}
\mathbf{Y}=\mathbf{\Phi} \cdot \mathbf{X},
\end{equation}
where $\mathbf{Y} \in R^M$ and $\mathbf{\Phi}$ is an $M\times N$ matrix.
The striking feature of CS is that the number of measurements can be
much less than the number of components of the unknown vector, i.e.,
$M\ll N$, insofar as $\mathbf{X}$ is sparse and the number of non-zero
components in it is less than $M$. Accurate reconstruction can be achieved by
solving the following convex-optimization problem~\cite{CS1}:
\begin{eqnarray}
\min \| \mathbf{X}\|_1  \ \ \hbox{subject to} \ \ \mathbf{Y} =
\mathbf{\Phi} \cdot \mathbf{X},
\end{eqnarray}
where $\| \mathbf{X}\|_1 = \sum_{i=1}^N | \mathbf{X}_i|$ is the $L_1$
norm of $\mathbf{X}$ and the matrix $\mathbf{\Phi}$ satisfies restricted
isometry property. Solutions to the convex optimization
are now standard~\cite{CS1,CS2,CS3,CS4,CS5,CS6}. (More details of the
CST can be found in Supplementary Note~1.) Our goal is to develop
a framework to cast the problem of reconstructing propagation
networks into the form~(\ref{eq:cs}).\\


\noindent
{\bf Reconstruction framework.}
To present our framework in a transparent manner, we first consider the
relatively simple case where there is no hidden source. Further, we
assume that the disease starts to propagate from a fraction of the infected
nodes. As we will see, based on this framework, it is feasible
to locate any hidden source based solely on time series after outbreak of
infection. The state of an arbitrary node $i$ is denoted as $S_i$, where
\begin{eqnarray}
S_i=\left\{
      \begin{array}{ll}
        0, & \hbox{susceptible;} \\
        1, & \hbox{infected.}
      \end{array}
    \right.
\end{eqnarray}
Due to the characteristic difference between the SIS dynamics and CP,
we treat them separately (see Methods).

For the SIS dynamics, the probability $P_i^{01}(t)$ of an
arbitrary node $i$ being infected by its neighbors at time $t$ is
\begin{equation} \label{eq:SIS_infect}
P_{i}^{01}(t) =1-(1-\lambda_{i})^{\sum_{j=1,j\ne i}^{N} a_{ij}S_{j}(t)},
\end{equation}
where $\lambda_i$ is the infection rate of $i$, $a_{ij}$ stands for the
elements of the adjacency matrix ($a_{ij}=1$ if $i$ connects to $j$ and
$a_{ij}=0$ otherwise), $S_j(t)$ is the state of node $j$ at
$t$, and the superscript $01$ denotes the change from susceptible state
(0) to infected state (1). At the same time, the recovery probability
of $i$ is $P_{i}^{10}(t)=\delta_{i}$,
where $\delta_i$ is the recovery rate of node $i$ and the superscript $10$
denotes the transition from infected state to susceptible state.
Equation~(\ref{eq:SIS_infect}) can be rewritten as
\begin{equation}
\label{eq:sis02}
\ln[1-P_{i}^{01}(t)]=\ln(1-\lambda_{i}) \cdot \sum_{j=1,j\ne i}^{N} a_{ij}S_{j}(t).
\end{equation}
If measurements at different times $t= t_1, t_2, \cdots, t_m$ are
available, equation~(\ref{eq:sis02}) can be written in the matrix form
$\mathbf{Y}_{m\times 1} = \mathbf{\Phi}_{m\times (N-1)}\cdot
\mathbf{X}_{(N-1)\times 1}$, where $\mathbf{Y}$ contains
$\ln[1-P_{i}^{01}(t)]$ at different $t$, $\mathbf{\Phi}$ is determined by
the state $S_j(t)$ of nodes except $i$, and $\mathbf{X}$ comprising the links and
infection rates of $i$ is sparse for a general network (see Methods). The main
challenge here is that the infection probabilities $P_i^{01}(t)$ at different
times are not given directly from the time series of the nodal state.

To develop a method to estimate the probability from the nodal states,
we set a threshold $\Delta$ pertaining to the normalized Hamming
distance between strings composed of $S_j(t)$ ($j\neq i$) at different $t$
to identify a base string at $\hat{t}_\alpha$ and a set of strings subject to the base.
According to the law of large numbers, the probability $P_{i}^{01}(\hat{t}_\alpha)$
can be estimated by the average over the state $S_i(t+1)$ at all proper time.
By setting another threshold $\Theta$ associated with the normalized Hamming distance,
we can identify a set of base strings. This process finally gives rise to
a set of reconstruction equations in the matrix form:
\begin{center}\footnotesize
\begin{eqnarray}
\label{eq:SIS_matrix}
\left[
\begin{array}{c}
\ln[1-\langle S_i(\hat t_1+1)\rangle]\\
\ln[1-\langle S_i(\hat t_2+1)\rangle] \\
\vdots\\
\ln[1-\langle S_i(\hat t_m+1)\rangle]
\end{array}
\right]=
\left[
\begin{array}{cccccc}
\langle S_{1}(\hat t_{1})\rangle  &\cdots& \langle S_{i-1}(\hat t_{1})\rangle
& \langle S_{i+1}(\hat t_{1})\rangle &\cdots& \langle S_{N}(\hat t_{1})\rangle \\
\langle S_{1}(\hat t_{2})\rangle  &\cdots& \langle S_{i-1}(\hat t_{2})\rangle
& \langle S_{i+1}(\hat t_{2})\rangle &\cdots& \langle S_{N}(\hat t_{2})\rangle \\
\vdots &\vdots & \vdots &\vdots & \vdots & \vdots\\
\langle S_{1}(\hat t_{m})\rangle  &\cdots& \langle S_{i-1}(\hat t_{m})\rangle
& \langle S_{i+1}(\hat t_{m})\rangle &\cdots& \langle S_{N}(\hat t_{m})\rangle
\end{array}
\right]
\left[
\begin{array}{c}
\ln(1-\lambda_i)a_{i1}\\
\vdots\\
\ln(1-\lambda_i)a_{i,i-1}\\
\ln(1-\lambda_i)a_{i,i+1}\\
\vdots\\
\ln(1-\lambda_i)a_{iN}
\end{array}
\right],
\end{eqnarray}
\end{center}
%
where $\hat t_1, \hat t_2,\cdots , \hat t_m$ correspond to the time associated
with $m$ base strings and $\langle \cdot \rangle$ denote the average over all
satisfied $t$ (see Methods). The vector $\mathbf{Y}_{m\times 1}$ and the matrix
$\mathbf{\Phi}_{m\times (N-1)}$ can then be obtained based solely on
time series of nodal states and the vector $\mathbf{X}_{(N-1)\times 1}$
to be reconstructed is sparse, rendering applicable the CS framework. As
a result, we can achieve exact reconstruction of all neighbors of node $i$
from relatively small amounts of observation. In a similar fashion the
neighboring vectors of all other nodes can be uncovered from time series,
enabling a full reconstruction of the whole network by matching the
neighboring sets of all nodes.

\begin{figure}
\begin{center}
\epsfig{figure=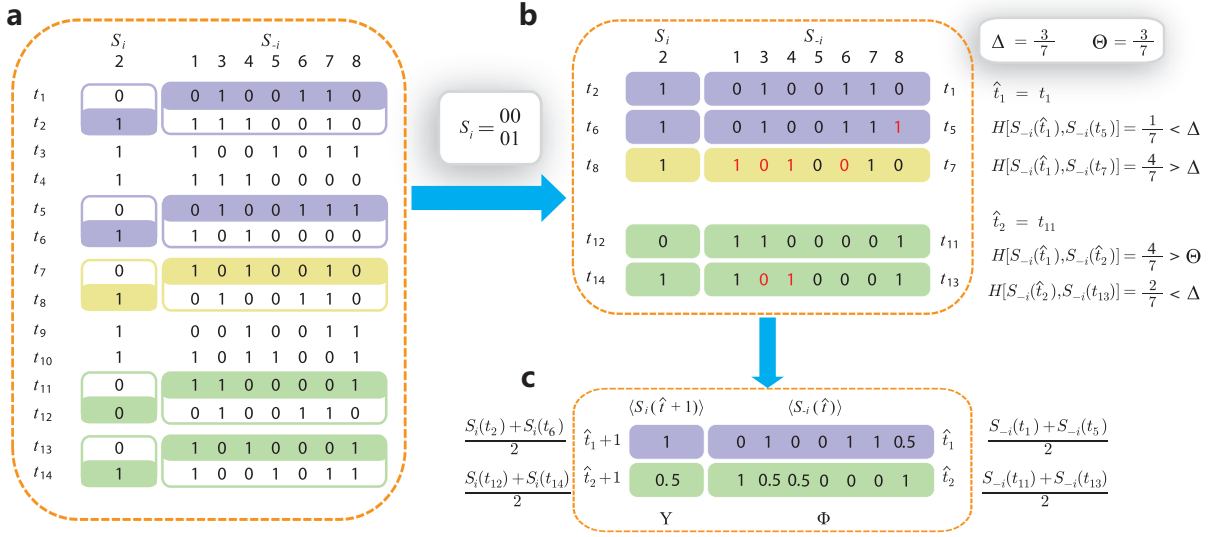,width=\linewidth}
\caption{\footnotesize{{\bf Schematic illustration of building up
$\mathbf{Y}$ and $\mathbf{\Phi}$ from binary time series.}
({\bf a}) 14 snapshots of data at the time instants $t_1$-$t_{14}$ of 8
nodes in a sample network, where $S_i$ is the time series of node 2 and
$S_{-i}$ denotes the strings of other nodes at different times. The
neighborhood of node 2 is to be reconstructed. Only the pairs 00 and 01
in the time series of $S_i$ ($i=2$) and the corresponding $S_{-i}$
contain useful information about the network,
as marked by different colors. ({\bf b}) Since $S_i(t+1)$ is determined by
the neighbors of $i$ and $S_{-i}(t)$, we sort out $S_i(t+1)$ and $S_{-i}(t)$
in the colored sections of the time series in ({\bf a}). According to the threshold
parameters $\Delta=3/7$ and $\Theta = 3/7$, we calculate the normalized
Hamming distance between each pair of strings $S_{-i}(t)$, finding
two base strings at $\hat{t}_1 = t_1$ and $\hat{t}_2 = t_{11}$ with
$H[S_{-i}(\hat{t}_1),S_{-i}(\hat{t}_2)]>\Theta$. We separate the colored strings
into two groups that are led by the two base strings, respectively.
In each group, the normalized Hamming distance
$H[S_{-i}(\hat{t}_\alpha),S_{-i}(t_\nu)]$
between the base string and other strings is calculated
and the difference from $S_{-i}(\hat{t}_\alpha)$ in each string is marked by red.
Using parameter $\Delta$, in the group led by $S_{-i}(\hat{t}_1)$,
$S_{-i}(t_5)$ and $S_i(t_6)$ are preserved, because of
$H[S_{-i}(\hat{t}_1),S_{-i}(t_5)]<\Delta$. In contrast, $S_{-i}(t_7)$
and $S_i(t_8)$ are disregarded because
$H[S_{-i}(\hat{t}_1),S_{-i}(t_7)]>\Delta$. In the group led by $S_{-i}(\hat{t}_2)$,
due to $H[S_{-i}(\hat{t}_2),S_{-i}(t_{13})]<\Delta$, the string is preserved.
The two sets of remaining strings marked by
purple and green can be used to yield the quantities required by the
reconstruction formula. (Note that different base strings are allowed to
share some strings, but for simplicity, this situation is not illustrated
here. See Supplementary Note~3 for a detailed discussion.) ({\bf c})
The average values $\langle S_i(\hat{t}_{\alpha}+1) \rangle$
and $\langle S_{-i}(\hat{t}_{\alpha})\rangle$ used to extract the vector $\mathbf{Y}$
and the matrix $\mathbf{\Phi}$ in the reconstruction formula, where
$\langle S_{-i}(\hat{t}_1)\rangle = [S_{-i}(t_1) + S_{-i}(t_5)]/2$,
$\langle S_{-i}(\hat{t}_2)\rangle = [S_{-i}(t_{11}) + S_{-i}(t_{13})]/2$,
$\langle S_i(\hat{t}_1+1)\rangle = [S_i(t_2) + S_i(t_6)]/2$,
and $\langle S_i(\hat{t}_2+1)\rangle = [S_i(t_{12}) + S_i(t_{14})]/2$
based on the remaining strings marked in different colors
(see Methods for more details). CST can be used to reconstruct the neighboring
vector $\mathbf{X}$ of node 2 from $\mathbf{Y}$ and $\mathbf{\Phi}$
from $\mathbf{Y}$= $\mathbf{\Phi} \cdot \mathbf{X}$.} }
\label{fig:Hamming}
\end{center}
\end{figure}

For the CP dynamics, the infection probability of an arbitrary
node $i$ is given by
\begin{eqnarray}
\label{eq:cp01}
P_{i}^{01}(t)=\lambda_{i}\sum_{j=1,j\ne i}^{N} a_{ij}S_{j}(t)/k_{i},
\end{eqnarray}
where $k_i$ is the degree of the node $i$, and the recovery probability is
$P_{i}^{10}(t)=\delta_{i}$ (see Methods). In close analogy to the SIS dynamics,
we have
\begin{equation}
\langle  S_{i}(\hat t_{\alpha}+1)\rangle  \simeq\langle  P_{i}^{01}(\hat t_{\alpha})\rangle
=\frac{\lambda_{i}\sum a_{ij}\langle  S_{j}(\hat t_{\alpha})\rangle}{ k_{i}}.
\label{eq:cp02}
\end{equation}
We then choose a series of base strings using a proper threshold $\Theta$
to establish a set of equations, expressed in the matrix form
$\mathbf{Y}_{m\times 1} = \mathbf{\Phi}_{m\times (N-1)} \cdot
\mathbf{X}_{(N-1)\times 1}$ (see Supplementary Note~2), where $\mathbf{\Phi}$
has the same form as in equation~(\ref{eq:SIS_matrix}), but $\mathbf{Y}$ and
$\mathbf{X}$ are given by
\begin{eqnarray}
\mathbf{Y} &=& \left[\langle S_i(\hat{t}_1 +1)\rangle, \langle S_i(\hat{t}_2 +1)\rangle,
\cdots, \langle S_i(\hat{t}_m +1)\rangle  \right]^\text{T},  \nonumber \\
\mathbf{X} &=& \left[ \frac{\lambda_i}{k_i}a_{i1}, \cdots,
\frac{\lambda_i}{k_i}a_{i,i-1}, \frac{\lambda_i}{k_i}a_{i,i+1}, \cdots,
\frac{\lambda_i}{k_i}a_{iN}  \right]^\text{T}.
\end{eqnarray}
Our reconstruction framework based on establishing the vector $\mathbf{Y}$ and
the matrix $\mathbf{\Phi}$ is schematically illustrated in Fig.~\ref{fig:Hamming}.
It is noteworthy that our framework can be extended to directed networks
in a straightforward fashion due to the feature that the neighboring set of
each node can be independently reconstructed. For instance, the neighboring
vector $\mathbf{X}$ can be defined to represent a unique link direction, e.g.,
incoming links. Inference of the directed links of all nodes yields
the full topology of the entire directed network. \\

\noindent
{\bf Reconstructing networks and inhomogeneous infection and recovery rates.}
To quantify the performance of our method in terms of the number of
base strings (equations) for a variety of diffusion dynamics and network
structures, we study the success rates for existent links (SREL) and null
connections (SRNC), corresponding to non-zero and zero element values in
the adjacency matrix, respectively. We impose the strict criterion that,
the network is regarded to have been fully reconstructed if and only if
both success rates reach 100$\%$. The sparsity of links makes it necessary to
define SREL and SRNC separately. Since the reconstruction method is
implemented for each node in the network, we define SREL and SRNC on the basis
of each individual node and, the two success rates for the entire network
are the respective averaged values over all nodes.
We also consider the issue of trade-off in terms of the
true positive rate (TPR - for correctly inferred links) and the false
positive rate (FPR - for incorrectly inferred links).

\begin{figure}
\begin{center}
\epsfig{figure=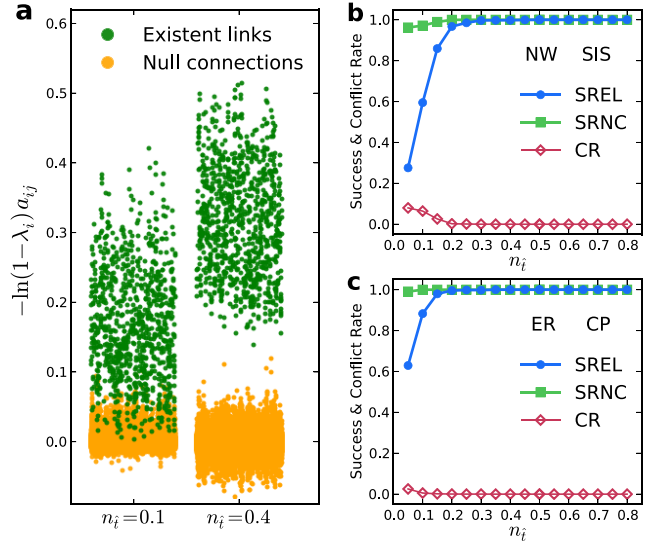,width=85mm}
\caption{{\bf Network reconstruction performance.}
({\bf a}) Element values $\ln(1-\lambda_i)a_{ij}$ of vector $\mathbf{X}$
times -1 for different fraction $n_{\hat{t}}$ of base strings for SIS dynamics.
({\bf b}-{\bf c}), success rate (SREL and SRNC) and conflict rate (CR) of
reconstruction as a function of $n_{\hat{t}}$
for SIS dynamics on Newman-Watts (NW) small-world networks
({\bf b}) and CP dynamics on Erd\"os-R\'enyi (ER) random networks ({\bf c}),
respectively. For the SIS dynamics, the parameters are $\Theta=0.25$, $\Delta=0.45$,
the infection and recovery rates $\lambda_i$ and $\delta_i$ are randomly
distributed in the ranges $(0.2,0.4)$ and $(0.4,0.6)$, respectively.
For the CP dynamics, the parameters are $\Theta=0.35$, $\Delta=0.45$,
$\lambda_i$ and $\delta_i$ are randomly distributed in the ranges $(0.7,0.9)$
and $(0.2,0.4)$, respectively. The network size $N$
is 200 with average node degree $\langle k\rangle =4$. The results are
obtained by ensemble averaging over 10 independent realizations.
The success rate is determined by setting a cut-off according to
Supplementary Fig.~1a and the method described in Supplementary Note~4.}
\label{fig:gap}
\end{center}
\end{figure}

Here, we assume that there is no hidden source and the spreading process starts
from a fraction of infected nodes, and record the binary time series.
Figure~\ref{fig:gap}(a) shows the reconstructed values of the components
of the neighboring vector $\mathbf{X}$ of all nodes. Let $n_{\hat{t}}$ be the
number of base strings normalized by the network size $N$. For small
values of $n_{\hat{t}}$, e.g., $n_{\hat{t}}=0.1$, the values of
elements associated with links and that associated with null connections
(actual zeros in the adjacency matrix) overlap, leading to ambiguity in
the identification of links. In contrast, for larger values of $n_{\hat{t}}$,
e.g., $n_{\hat{t}}=0.4$, an explicit gap emerges between the two groups
of element values, enabling us to correctly identify all links by simply
setting a cut-off within the gap (see Supplementary Fig.~1a and Note~4
for the method to set the cut-off). The success rates (SREL and SRNC) as
a function of $n_{\hat{t}}$ for SIS and CP on both homogeneous
and heterogeneous networks are shown in Figs.~\ref{fig:gap}(b,c), where
we observe nearly perfect reconstruction of links insofar as $n_{\hat{t}}$
exceeds a relatively small value - an advantage of compressed sensing. The
exact reconstruction is robust in the sense that a wide range of $n_{\hat{t}}$
values can yield nearly $100\%$ success rates. Our reconstruction method
is then effective for tackling real networks in the absence of any
a priori knowledge about its topology. In particular, the existence of
a clear gap in the reconstructed vector $\mathbf{X}$ represents a
successful reconstruction for a real network.

Note that a network is reconstructed through the union of all neighborhoods,
which may encounter ``conflicts'' with respect to presence/absence of a link between
two nodes as generated by reconstruction centered at the two nodes, respectively.
The conflicts would reduce the accuracy in the reconstruction of the entire
network. To characterize the effects of edge conflicts, we study the
consistency of mutual assessment of the presence or absence of link between
each pair of nodes, as shown in Figs.~\ref{fig:gap}(b,c). We see that inconsistency
arises for small values of $n_{\hat{t}}$ but vanishes completely when
the success rates reach $100\%$, indicating complete consistency among
the mutual inferences of nodes and consequently guaranteeing accurate
reconstruction of the entire network. Detailed results of success rates
and trade-off measures with respect to
a variety of model and real networks are displayed in Table I, Supplementary
Figs.~2 and 3 and Supplementary Note~5.

\begin{figure}
\begin{center}
\epsfig{figure=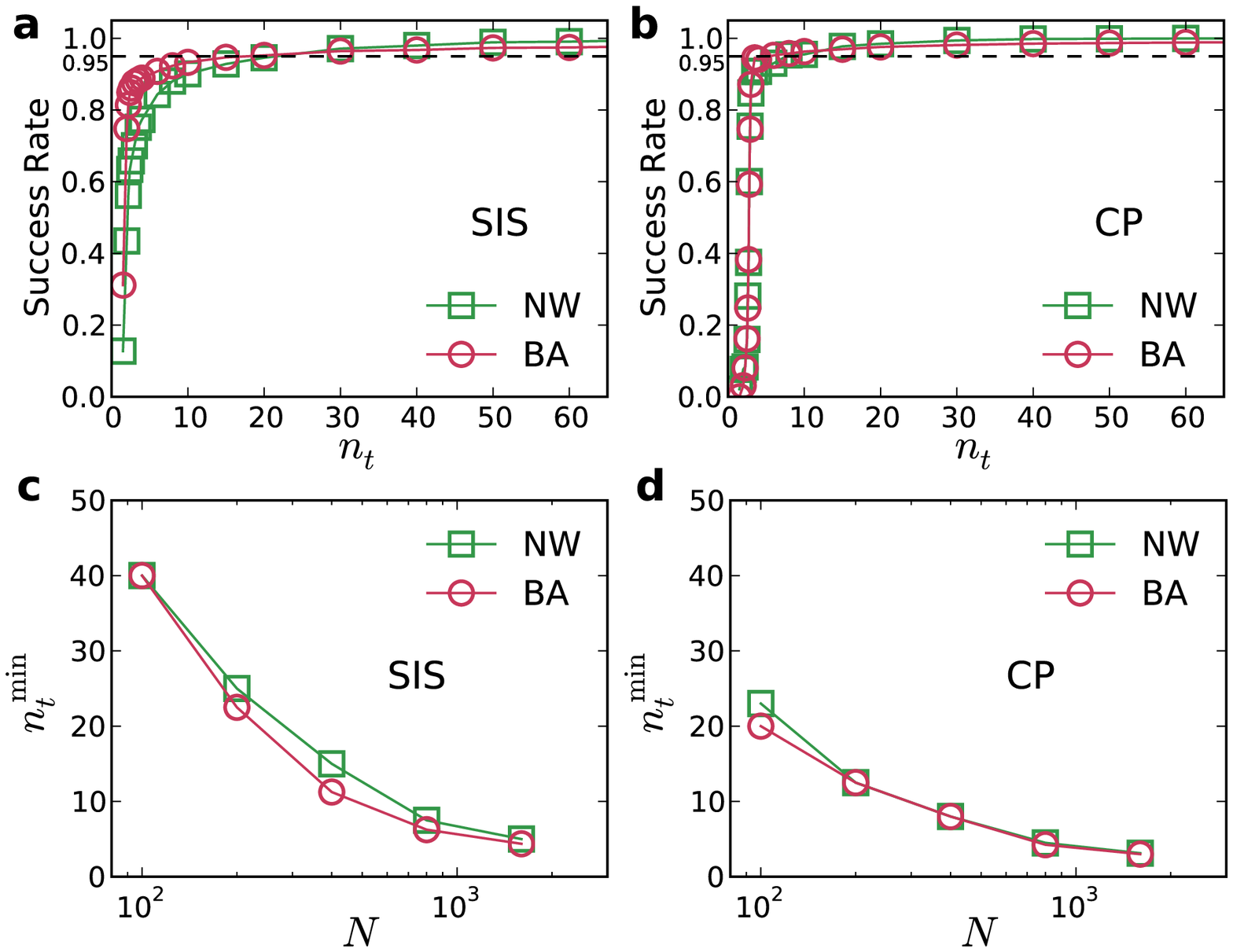,width=85mm}
\caption{{\bf Effect of the length of time series and network size.}
({\bf a}-{\bf b}) Success rate as a function of the relative length $n_t$ of
time series for SIS ({\bf a}) and CP ({\bf b}) in combination with NW and
Barab\'asi-Albert (BA) networks. The dashed lines represent 95$\%$
success rate. ({\bf c}-{\bf d}) the minimum relative
length $n_t^{\min}$ that assures at least $95\%$ success rate as a function
of the network size $N$ for SIS ({\bf c}) and CP ({\bf d}) dynamics
on NW and BA networks. Here, the success rate is the geometric average
over SREL and SRNC. For SIS dynamics, $\lambda_i \in (0.1,0.3)$ and
$\delta \in (0.2,0.4)$. For CP dynamics, $\lambda_i \in (0.7,0.9)$ and
$\delta \in (0.3,0.5)$. In ({\bf a}) and ({\bf b}), the network size $N$ is 500.
The other parameters are the same as in Fig.~\ref{fig:gap}. Note that
$n_t \equiv t/N$, where $t$ is the absolute length of time series, and
$n_t^{\min} \equiv t_{\min}/N$, where $t_{\min}$ is the minimum absolute
length of time series required for at least $95\%$ success rate.}
\label{fig:length}
\end{center}
\end{figure}

Although the number of base strings is relatively small compared to
the network size, we need a set of strings at different time with
respect to a base string to formulate the mathematical framework for
reconstruction. We study how the length of time series affects the
accuracy of reconstruction. Figures~\ref{fig:length}(a,b) show the
success rate as a function of the relative
length $n_t$ of time series for SIS and CP dynamics on both homogeneous
and heterogeneous networks, where $n_t$ is the total length of time series
from the beginning of the spreading process divided by the network size $N$.
The results demonstrate that even for very small values of $n_t$, most
links can already be identified, as reflected by the high values of the
success rate shown. Figures~\ref{fig:length}(c,d) show the minimum length
$n_t^\text{min}$ required to achieve at least $95\%$ success rate for
different network sizes. For both SIS and CP dynamics on different networks,
$n_t^\text{min}$ decreases considerably as $N$ is increased. This
seemingly counterintuitive result is due to the fact that different base
strings can share strings at different times to enable reconstruction.
In general, as $N$ is increased, $n_{\hat{t}}$ will increase accordingly.
However, a particular string can belong to different base strings with
respect to the threshold $\Delta$, accounting for the slight increase
in the absolute length of the time series (see Supplementary Fig.~4 and
Note~5) and the reduction in $n_t^\text{min}$
(see Supplementary Note~3 on the method to choose base and subordinate
strings). The dependence of the success rate on the average node degree
$\langle k\rangle$ for SIS and CP on different networks has been
investigated as well (see Supplementary Fig.~5 and Note~5). The results
in Fig.~\ref{fig:gap}, Fig.~\ref{fig:length}, Supplementary Fig.~2-5
and Table 1 demonstrate the high accuracy and efficiency of our reconstruction
method based on small amounts of data.

\begin{figure}
\begin{center}
\epsfig{figure=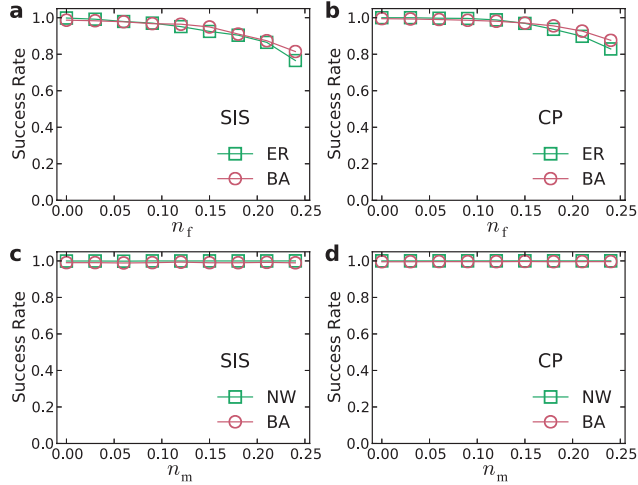,width=85mm}
\caption{{\bf Reconstruction against noise and inaccessible nodes.}
({\bf a}-{\bf b}) Success rate as a function of the fraction $n_\text{f}$ of
the flipped states induced by noise in time series for SIS ({\bf a}) and CP ({\bf b})
dynamics on ER and BA networks. ({\bf c}-{\bf d}) success rate as a function of
the fraction $n_\text{m}$ of nodes that are externally inaccessible
for SIS ({\bf c}) and CP ({\bf d}) dynamics on
NW and BA networks. The network size $N$ is 500 and $\langle k\rangle=4$.
The other parameters are the same as in Fig.~\ref{fig:gap}.}
\label{fig:noise}
\end{center}
\end{figure}

In practice, noise is present and it is also common for time series from
certain nodes to be missing, and it is necessary to test the applicability
of our method in more realistic situations. Figures~\ref{fig:noise}(a,b) show the
dependence of the success rate on the fraction $n_\text{f}$ of states in the
time series that flip due to noise for SIS and CP dynamics on two types of
networks. We observe that the success rates are hardly affected, providing strong
evidence for the applicability of our reconstruction method. For example,
even when $25\%$ of the nodal states flip, we can still achieve about $80\%$
success rates for both dynamical processes and different network topologies.
Figures~\ref{fig:noise}(c,d) present the success rate versus the fraction
$n_\text{m}$ of unobservable nodes, the states of which are externally inaccessible.
We find that the high success rate remains mostly unchanged as $n_\text{m}$
is increased from zero to $25\%$, a somewhat counterintuitive but striking
result. The high degree of robustness against the limit to access nodal
states is elaborated further in Supplementary Fig.~6 and Note~5. We find
that, in general, missing information can affect the reconstruction of the
neighboring vector, as reflected by the reduction of the gap between the
reconstructed values associated with actual links and null connections. However,
even for high values of $n_\text{m}$, e.g., $n_\text{m}=0.3$, there is
still a clear gap, indicating that a full recovery of all links is achievable.
We have also found that our method is robust against inaccurately
specified diffusion processes with fluctuation in infection rates
(see Supplementary Fig.~7 and Note~5).
Taken together, the high accuracy, efficiency and robustness against noise,
missing information and inaccurately modeling real dynamical processes
provide strong credence for the validity and power
of our framework for binary time-series based network reconstruction.


Having reconstructed the network structure, we can estimate the infection
and recovery rates of individuals to uncover their diversity in immunity.
This is an essential step to implement target vaccination strategy in a
population or on a computer network to effectively suppress/prevent the
spreading of virus at low cost, as a large body of literature indicates
that knowledge about the network structure and individual characteristics
is sufficient for controlling the spreading dynamics~\cite{PhysRevLett.91.247901,
Forster20032007,Klepac23082011,Kleczkowski07012012}.
Here we offer an effective method to infer the individuals' infection rates
$\lambda_i$ based solely on the binary time series of the nodal states after
an outbreak of contamination. (To our knowledge, there was no prior work
addressing this critical issue.) In particular, after all links have been
successfully predicted, $\lambda_i$ can be deduced from the infection
probabilities that can be approximated by the corresponding infection
frequencies (see Methods). These probabilities depend on both $\lambda_i$
and the number of infected neighbors. The reproduced infection
rates $\lambda_i$ of individuals for both SIS and CP dynamics on
different networks are in quite good agreement with the true values
with small prediction errors (see Supplementary Fig.~8 and Note~6).
Results from a comprehensive error analysis are listed in Table I,
where the uniformly high accuracy validates our method. The inhomogeneous
recovery rates $\delta_i$ of nodes can be predicted from the binary time
series in a more straightforward way, because $\delta_i$'s do not depend on
the nodal connections (see Supplementary Fig.~9 and Note~6). Thus our
framework is capable of predicting characteristics of nodal diversity
in terms of degrees and infection and recovery rates based solely on
binary time series of nodal states. \\


\noindent
{\bf Locating the hidden source of propagation.} We assume that a hidden source
exists outside the network but there are connections between it and some
nodes in the network. In practice, the source can be modeled as a special
node that is always infected. Starting from the neighborhood of the source,
the infection originates from the source and spreads all over the network.
We collect a set of time series of the nodal states except the hidden source
(see Methods). The basic idea of ascertaining and locating the hidden source
is based on missing information from the hidden source when attempting to
reconstruct the network. In particular, in order to reconstruct the
connections belonging to the immediate neighborhood of the source
accurately, time series from the source are needed to generate the
matrix $\mathbf{\Phi}$ and the vector $\mathbf{Y}$.
But since the source is hidden, no time series from it are available, leading
to reconstruction inaccuracy and, consequently, anomalies in the
predicted link patterns of the neighboring nodes. It is then possible to
detect the neighborhood of the hidden source by identifying any abnormal
connection patterns~\cite{SWL:2012}, which can
be accomplished by using different data segments.
If the inferred links of a node are stable with
respect to different data segments, the node can be deemed to have no
connection with the hidden source; otherwise, if the result of inferring a
node's links varies significantly with respect to different data segments,
the node is likely to be connected to the hidden
source. The standard deviation of the predicted results with respect
to different data segments can be used as a quantitative criterion for
the anomaly. Once the neighboring set of the source is determined, the
source is then precisely located topologically.

\begin{figure}
\begin{center}
\epsfig{figure=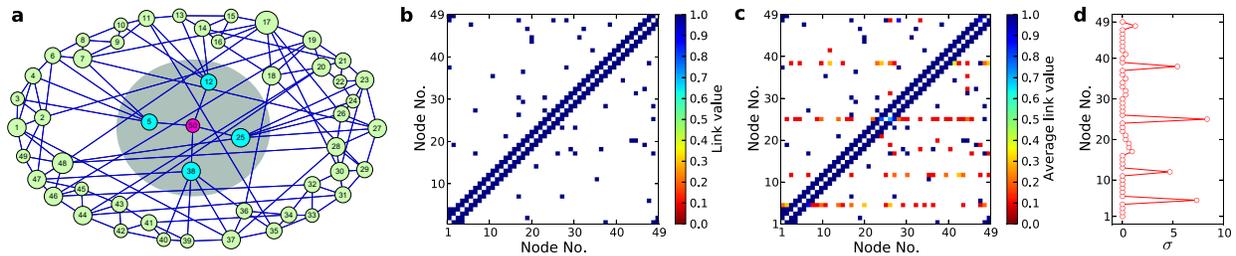,width=\linewidth}
\caption{{\bf Locating an external hidden source
from time series.} ({\bf a}) Hidden
source treated as a special node (in red) is connected to four nodes in the
network (blue). The time series of other nodes except the source (No. 50)
after the outbreak of an epidemic are assumed to be available. ({\bf b}) True
adjacency matrix of the NW network with identical link weights
to facilitate a comparison with the reconstructed adjacency matrix.
({\bf c}) Reconstructed adjacency matrix from a number of segments in time series. The
four neighboring nodes of the source are predicted to be densely linked to
other nodes, as indicated by the average value of the elements in the four rows
corresponding to these nodes. ({\bf d}) Structural variance $\sigma$ of each node.
The four neighboring nodes of the source exhibit much larger values of $\sigma$
than those from the other nodes, providing unequivocal evidence that they belong
to the immediate neighborhood of the hidden source.}
\label{fig:hidden}
\end{center}
\end{figure}

Figure~\ref{fig:hidden} presents an example,
where a hidden source is connected with four nodes in the network
[Fig.~\ref{fig:hidden}(a)], as reflected in the network adjacency matrix
[Fig.~\ref{fig:hidden}(b)]. We implement our reconstruction framework on
each accessible node by using different sets of data in the time series. For each data
set, we predict the neighbors of all nodes and generate an adjacency matrix.
Averaging over the elements corresponding to each location in all the
reconstructed adjacency matrices, we obtain Fig.~\ref{fig:hidden}(c), in
which each row corresponds to the mean number of links in a node's neighborhood.
The inferred links of the immediate neighbors of the hidden source
exhibit anomalies. To quantify the anomalies, we
calculate the structural standard deviation $\sigma$ from
different data segments, where $\sigma$ associated with node $i$ is
defined through the $i$th row in the adjacency matrix as
\begin{equation} \label{eq:struc_var}
\sigma_i = \frac{1}{N}\sum_{j=1}^{N}\sqrt{\frac{1}{g}
\sum_{k=1}^{g}\big(a_{ij}^{(k)} - \langle a_{ij}  \rangle   \big)^2},
\end{equation}
where $j$ denotes the column, $a_{ij}^{(k)}$ represents the element value in
the adjacency matrix inferred from the $k$th group of the data,
$\langle a_{ij}\rangle = (1/g)\sum_{k=1}^{g} a_{ij}^k$ is the mean
value of $a_{ij}$, and $g$ is the number of data segments. Applying
equation~(\ref{eq:struc_var}) to the reconstructed adjacency matrices gives
the results in Fig.~\ref{fig:hidden}(d), where the values of $\sigma$
associated with the immediate neighboring nodes of the hidden source are
much larger than those from others (which are essentially zero). A cut-off value
can be set in the distribution of $\sigma_i$ to identify the immediate
neighbors of the hidden source (see Supplementary Fig.~1b and Note~4).
The performance of locating hidden source by means of the trade-off measures
(TPR versus FPR) are displayed in Table I.\\

\noindent
{\bf \large{Discussion}} \\
We have developed a general framework to reconstruct complex propagation
networks upon which epidemic spreading takes place from binary time series.
Our paradigm is based on compressed sensing, completely data-driven,
and practically significant for controlling the epidemic spreading
through targeted vaccination. Both theoretically and practically,
our framework can be used to address the extremely challenging problem
of reconstructing the intrinsic interacting patterns of complex stochastic
systems based on small amounts of polarized time series.
The key to success of our method lies in our development of a novel
class of transformation allowing the network inference problem to be
converted to the problem of sparse signal reconstruction, which can then
be solved by the standard compressed-sensing algorithm. The accuracy
and efficiency of our framework in uncovering the network structure, the
natural diversity in the nodal characteristics, and any hidden source
are guaranteed by the compressed sensing theory with rigorous proof for
low-data requirement and convergence to optimal solution. The feasibility
of our framework has been demonstrated using a large number of combinations of
epidemic processes and network structures, where in all cases
extremely accurate reconstruction is achieved.
Our approach opens up a new avenue towards fully addressing the inverse
problem in complex stochastic systems in a highly
efficient manner, a fundamental stepping stone toward understanding
and controlling complex dynamical systems in general.

We have focused on two types of spreading dynamics, SIS and CP, where
an infected individual can recover and becomes susceptible again. In
this regard, even if an outbreak occurs, control strategy such as targeted
vaccination or quarantine can be helpful to eliminate the virus
eventually. A main purpose of our work is to identify the key individuals
in the network to implement target control and to locate the source of
infection to isolate it so as to prevent recurrent infection in the future.
While for any spreading dynamics, the most effective way to prevent
a large-scale outbreak is to implement control during the early stage,
this may be impractical in many situations. If we miss the
early stage, which is possible especially in complex networks where the
epidemic threshold can be near zero, to be able to reconstruct the
spreading network is of tremendous value. Besides disease
spreading, our framework is applicable to rumor or information
spreading. In this case, identifying the source of rumor is important,
a problem that our framework is capable of solving.

Our work raises a number of questions to further and perfect
the theoretical and algorithmic development in reconstructing complex
dynamical systems. For example, if partial knowledge about the network
structure is available, the information can be incorporated into our
framework to further reduce the required data amount. Moreover,
for non-Markovian spreading processes, our current reconstruction
framework may fail. This raises the need to develop new and more
general approaches. Nevertheless, our theory, due
to its generality and applicability to various types of inhomogeneous
interactions, can be applied to networks of networks or interdependent
networks, in which there may be different spreading patterns associated
with distinct layers or components. Taken together, our results provide
strong credence to the proposition that complex networks can be fully
decrypted from measurements, even when stochastic disturbance
and hidden sources are present. This can offer a deeper understanding of
complex systems in general and significantly enhance our ability to
control them based on, e.g., the recently developed controllability
theory of complex networks~\cite{Slotine:book,liu:2011,vicsek:2012,
Yan:2012,Liu:PNAS,PowerCont,Yuan:control}. \\

\noindent
{\bf \large{Methods}} \\
{\small
{\bf Spreading processes.}
The SIS model is the classic epidemic model that has been used frequently to study
a variety of spreading behaviors in social and computer networks. Each node of
the network represents an individual and links are connections along which
the infection can propagate to others with certain probability.
At each time step, a susceptible node $i$ in state $0$
is infected with rate $\lambda_i$ if it is connected to an infected
node in state 1. If $i$ connects to more than one infected neighbor,
the infection probability $P^{01}$ is given by equation~(\ref{eq:SIS_infect}).
At the same time, infected nodes are continuously recovered to be susceptible at the
rates $\delta_i$. The CP model has been used extensively to
describe, e.g., the spreading of infection and competition of animals over a
territory, where $\lambda_i$ is determined by equation~(\ref{eq:cp01}).
The main difference between SIS and CP dynamics lies in the
influence on a node's state from its vicinity. In both SIS and CP
dynamics, $\lambda_i$ and $\delta_i$ depend on the individuals'
immune systems and are selected from a Gaussian
distribution characterizing the natural diversity (see Supplementary Note~7
for details of numerical simulations). Moreover, a hidden source is regarded
as infected for all time.\\




\noindent
{\bf Mathematical formulation of reconstruction based on CST.}
For SIS dynamics, suppose measurements at a
sequence of times $t= t_1, t_2, \cdots, t_m$ are
available. Equation~(\ref{eq:sis02}) leads to the following matrix form
$\mathbf{Y}_{m\times 1} = \mathbf{\Phi}_{m\times (N-1)}\cdot
\mathbf{X}_{(N-1)\times 1}$:
\begin{center} 
$\begin{bmatrix}
\ln[1-P_{i}^{01}(t_1)]\\
\ln[1-P_{i}^{01}(t_2)] \\
\vdots\\
\ln[1-P_{i}^{01}(t_m)]
 \end{bmatrix}$
=
$\begin{bmatrix}
S_{1}(t_{1})  &\cdots& S_{i-1}(t_{1}) & S_{i+1}(t_{1}) &\cdots& S_{N}(t_{1}) \\
S_{1}(t_{2})  &\cdots& S_{i-1}(t_{2}) & S_{i+1}(t_{2}) &\cdots& S_{N}(t_{2}) \\
\vdots &\vdots & \vdots &\vdots & \vdots & \vdots\\
S_{1}(t_{m})  &\cdots& S_{i-1}(t_{m}) & S_{i+1}(t_{m}) &\cdots& S_{N}(t_{m})
 \end{bmatrix}
 $ 	$\begin{bmatrix}
\ln(1-\lambda_i)a_{i1}\\
\vdots\\
\ln(1-\lambda_i)a_{i,i-1}\\
\ln(1-\lambda_i)a_{i,i+1}\\
\vdots\\
\ln(1-\lambda_i)a_{iN}
 \end{bmatrix}$,
\end{center}
where the vector $\mathbf{X}_{(N-1)\times 1}$ contains all possible connections
between node $i$ and all other nodes, and it is sparse for a general complex
network. We see that, if the vector $\mathbf{Y}_{m\times 1}$ and the matrix
$\mathbf{\Phi}_{m\times (N-1)}$ can be constructed from time series,
$\mathbf{X}_{(N-1)\times 1}$ can then be solved by using CST. The main
challenge here is that the infection probabilities $P_i^{01}(t)$ at different
times are not given directly by the time series of the nodal states. To devise
a heuristic method to estimate the probabilities,
we assume that the neighboring set $\Gamma_i$ of the node $i$
is known. The number of such neighboring nodes is given by $k_i$, the degree
of node $i$, and their states at time $t$ can be denoted as
\begin{equation}
S_{\Gamma_i}(t) \equiv \{ S_1(t), S_2(t), \cdots, S_{k_i}(t) \}.
\end{equation}
In order to approximate the infection probability, we use $S_i(t)=0$ so that
at $t+1$, the node $i$ can be infected with certain probability. In contrast, if
$S_i(t)=1$, $S_i(t+1)$ is only related with the recovery probability $\delta_i$.
Hence, we focus on the $S_i(t)=0$ case to derive $P_i^{01}(t)$. If we can find
two time instants: $t_{1},t_{2}\in T$ ($T$ is the length of time series), such
that $S_{i}(t_{1})=0$ and $S_{i}(t_{2})=0$, we can then calculate the normalized
Hamming distance $H[S_{\Gamma_i}(t_{1}),S_{\Gamma_i}(t_{2})]$ between
$S_{\Gamma_i}(t_{1})$ and $S_{\Gamma_i}(t_{2})$, where the normalized Hamming
distance between two strings of equal length is defined as ratio of the number
of positions with different symbols between them and the length of string. If
$H[S_{\Gamma_i}(t_{1}),S_{\Gamma_i}(t_{2})]=0$, we can regard the states at the
next time step, $S_{i}(t_{1}+1)$ and $S_{i}(t_{2}+1)$, as i.i.d Bernoulli trials.
In this case, using the law of large numbers, we have
\begin{equation} \label{eq:neighbor}
\lim_{l\to\infty} {{1}\over{l}}\sum_{\nu=1}^{l}S_{i}(t_{\nu}+1)
\to P_{i}^{01}(\hat t_{\alpha}), \ \ \forall \ \ t_{\nu}, S_{i}(t_{\nu})=0,\ \
H[S_{\Gamma_i}(\hat t_{\alpha}),S_{\Gamma_i}(t_{\nu})]=0.
\end{equation}
A more intuitive understanding of equation~(\ref{eq:neighbor}) is that, if the states
of $i$'s neighbors are unchanged, the fraction of times of $i$ being infected by its
neighbors over the entire time period will approach the actual infection probability
$P_i^{01}$. Note, however, that the neighboring set of $i$ is unknown and to be
inferred. A strategy is then to artificially enlarge the neighboring set
$S_{\Gamma_i}(t)$ to include all nodes in the network except $i$. In particular,
we denote
\begin{equation} \label{eq:string}
S_{-i}(t)\equiv \{S_{1}(t),S_{2}(t),\dots,S_{i-1}(t),S_{i+1}(t),\dots,S_{N}(t)\}.
\end{equation}
If $H[S_{-i}(t_{1}),S_{-i}(t_{2})]=0$, the condition
$H[S_{\Gamma_i}(t_{1}),S_{\Gamma_i}(t_{2})]=0$ will be ensured. Consequently,
due to the nature of i.i.d Bernoulli trials, from the law of large numbers,
we have
\begin{equation*}
\lim_{l\to\infty} {{1}\over{l}}\sum_{\nu=1}^{l}S_{i}(t_{\nu}+1)\to P_{i}^{01}(\hat t_{\alpha}),
\ \ \forall \ \ t_{\nu}, S_{i}(t_{\nu})=0, \ \ H[S_{-i}(\hat t_{\alpha}), S_{-i}(t_{\nu})]=0.
\end{equation*}
Hence, the infection probability $P_{i}^{01}(\hat t_{\alpha})$ of
a node at $\hat t_{\alpha}$ can be evaluated by
averaging over its states associated with zero normalized Hamming distance between the
strings of other nodes at some time associated with $\hat t_{\alpha}$. In practice,
to find two strings with absolute zero normalized Hamming distance is unlikely.
We thus set a threshold $\Delta$ so as to pick the suitable strings to approximate
the law of large numbers, that is
\begin{equation} \label{eq:Delta_SIS}
{{1}\over{l}}\sum_{\nu=1}^{l\gg 1}S_{i}(t_{\nu}+1)\simeq  {{1}\over{l}}\sum_{\nu=1}^{l\gg 1}P_i^{01}(t_{\nu}),
\ \ \forall \ \ t_{\nu}, S_{i}(t_{\nu})=0, \ \ H[S_{-i}(\hat t_{\alpha}), S_{-i}(t_{\nu})]<\Delta,
\end{equation}
where $S_{-i}(\hat t_{\alpha})$ serves as a base for comparison with
$S_{-i}(t)$ at all other times and ${{1}\over{l}}\sum_{\nu=1}^{l\gg 1}P_i^{01}(t_{\nu}) \simeq
P_{i}^{01}(\hat t_{\alpha})$.
Since $H[S_{-i}(\hat{t}_\alpha),S_{-i}(t_\nu)]$ is not exactly zero,
there is a small difference between $P_{i}^{01}(\hat{t}_{\alpha})$
and $P_{i}^{01}(t_{\nu})$ ($\nu=1,\cdots,l$). We thus consider the
average of $P_{i}^{01}(t_{\nu})$ for all $t_\nu$ to obtain
$P_{i}^{01}(\hat t_{\alpha})$, leading to the right-hand side of
equation~(\ref{eq:Delta_SIS}). We denote $\langle S_i(\hat t_{\alpha}+1)\rangle
= {{1}\over{l}}\sum_{\nu=1}^{l\gg 1}S_{i}(t_{\nu}+1)$ and $\langle P_{i}^{01}(\hat t_{\alpha})
\rangle = {{1}\over{l}}\sum_{\nu=1}^{l\gg 1}P_i^{01}(t_\nu)$.
In order to reduce the error in the estimation, we implement the average on
$S_{-i}(t)$ over all selected strings through equation~(\ref{eq:Delta_SIS}).
The averaging process is with respect to the nodal states $S_{j,j\neq i}(t)$ on
the right-hand side of the modified dynamical equation (\ref{eq:sis02}). Specifically,
averaging over time $t$ restricted by equation~(\ref{eq:Delta_SIS}) on both sides of
equation~(\ref{eq:sis02}), we obtain
$\langle \ln[1-P_{i}^{01}(t)]\rangle =\ln(1-\lambda_{i})\sum_{j=1,j\ne i}^{N}
a_{ij}\langle S_{j}(t)\rangle$.
If $\lambda_i$ is small with insignificant fluctuations, we can approximately have
$\ln[1-\langle P_{i}^{01}(t)\rangle ]\simeq \langle \ln[1-P_{i}^{01}(t)]\rangle$
(see Supplementary Fig.~10 and Note~8),
which leads to
$\ln[1-\langle P_{i}^{01}(t)\rangle ]\simeq  \ln(1-\lambda_{i})
\sum_{j=1,j\ne i}^{N} a_{ij}\langle S_{j}(t)\rangle$.
Substituting $\langle P_{i}^{01}(\hat{t}_\alpha)\rangle$ by
$\langle  S_{i}(\hat t_{\alpha}+1)\rangle$, we finally get
\begin{eqnarray}
\ln[1-\langle S_{i}(\hat{t}_\alpha+1)\rangle ]\simeq \ln(1-\lambda_{i}) \cdot
\sum_{j=1,j\ne i}^{N} a_{ij}\langle S_{j}(\hat{t}_\alpha)\rangle.
\label{eq:intui_SIS}
\end{eqnarray}
While the above procedure yields an equation that bridges the links of
an arbitrary node $i$ with the observable states of the nodes, a single
equation does not contain sufficient structural information about the
network. Our second step is then to derive a sufficient number of
linearly independent equations required by CST to reconstruct the
local connection structure. To achieve this, we choose a series of
base strings at a number of time instants from a set denoted by
$T_{base}$, in which each pair of strings satisfy
\begin{equation}
H[S_{-i}(\hat t_{\beta}),S_{-i}(\hat t_{\alpha})]> \Theta,\quad
\forall \hat t_{\alpha},\hat t_{\beta} \in T_{base},
\end{equation}
where $\hat t_{\alpha}$ and $\hat t_{\beta}$ correspond to the time
instants of two base strings in the time series and $\Theta$ is a threshold.
For each string, we repeat the process of establishing the relationship
between the nodal states and connections, leading to a set of equations
at different values of $\hat t_{\alpha}$ in equation~(\ref{eq:intui_SIS}),
as described in the matrix form [equation~(\ref{eq:SIS_matrix})]. See
Supplementary Fig.~11, 12 and Note~8 for the dependence of success rate
on threshold $\Delta$ and $\Theta$ for SIS and CP dynamics in combination
with four types of networks. \\

\noindent
{\bf Inferring inhomogeneous infection rates.}
The values of the infection rate $\lambda_i$ of nodes can be inferred
after the neighborhood of each node has been successfully reconstructed.
The idea roots in the fact that the infection probability of a node
approximated by the frequency of being infected calculated from time
series is determined both by its infection rate and by the number of
infected nodes in its neighborhood. To provide an intuitive picture, we
consider the following simple scenario in which the number of infected
neighbors of node $i$ does not change with time. In this case, the
probability of $i$ being infected at each time step is fixed. We can
thus count the frequency of the $01$ and $00$ pairs embedded in the
time series of $i$. The ratio of the number of $01$ pairs over the
total number of $01$ and $00$ pairs gives approximately the infection
probability. The infection rate can then be calculated by using
equations ~(\ref{eq:SIS_infect}) and (\ref{eq:cp01})
for the SIS and CP dynamics, respectively. In a
real-world situation, however, the number of infected neighbors
varies with time. The time-varying factor can be taken into account
by sorting out the time instants corresponding to different numbers
of the infected neighbors, and the infection probability can be
obtained at the corresponding time instants, leading to a set of
values for the infection rate whose average represents an accurate
estimate of the true infection rate for each node.

To be concrete, considering all the time instants $t_{\nu}$ associated with $k_\text{I}$
infected neighbors, we denote $S_{i}^{(k_\text{I})}= (1/l)\sum_{\nu=1}^{l}S_{i}
(t_{\nu}+1)$, $\forall$ $t_{\nu}$, $\sum_{j\in\Gamma_{i}} S_{j}(t_{\nu}) = k_\text{I}$
and $S_i(t_\nu)=0$,
where $\Gamma_i$ is the neighboring set of node $i$, $k_\text{I}$ is the number of
infected neighbors, and $S_{i}^{(k_\text{I})}$ represents the average infected fraction
of node $i$ with $k_\text{I}$ infected neighbors. Given
$S_{i}^{(k_\text{I})}$, we can rewrite equation~(\ref{eq:SIS_infect})
by substituting $S_{i}^{(k_\text{I})}$ for $P_{i}^{01}(t)$
and $\lambda_i^{(k_\text{I})}$ for $\lambda_i$,
which yields $\lambda_{i}^{(k_\text{I})} = 1-\exp\left[\ln \big( 1
- S_{i}^{(k_\text{I})}\big) /k_I\right]$.
To reduce the estimation error, we average $\lambda_{i}^{(k_\text{I})}$ with respect
to different values of $k_\text{I}$, as follows:
\begin{equation} \label{eq:lambda_SIS}
\lambda_i^{\text{true}}(\text{SIS}) \approx \langle \lambda_i^{(k_\text{I})}  \rangle
= \frac{1}{N_{\Lambda_{i}}}\sum_{k_\text{I}\in \Lambda_{i}}\lambda_{i}^{(k_\text{I})},
\end{equation}
where $\Lambda_{i}$ denotes the set of all possible infected neighbors during
the epidemic process and $N_{\Lambda_{i}}$ denotes the number of different
values of $k_\text{I}$ in the set. Analogously, for CP, we can evaluate
$\lambda_i^{\text{true}}$ from equation~(\ref{eq:cp01}) by
\begin{equation} \label{eq:lambda_CP}
\lambda_i^{\text{true}}(\text{CP}) \approx \langle \lambda_i^{(k_\text{I})}  \rangle
= \frac{1}{N_{\Lambda_{i}}}\sum_{k_\text{I}\in \Lambda_i}
\frac{S_i^{(k_\text{I})} k_{i}}{k_{\text{I}}}
\end{equation}
where $k_i = \sum_{j=1}^{N} a_{ij}$ is the node degree of $i$. Insofar as
all the links of $i$ have been successfully reconstructed, $S_i^{(k_\text{I})}$ can be
obtained from the time series in terms of the satisfied $S_i(t_\nu +1)$, allowing
us to infer $\lambda_i^\text{true}$ via equation~(\ref{eq:lambda_SIS}) and
equation~(\ref{eq:lambda_CP}).

Note that the method is applicable to any type of networks insofar as the
network structure has been successfully reconstructed. \\

\noindent
{\bf Networks analysed.} Model networks and real networks we used are described
in Supplementary Note~10 and Table~1.}


\noindent
{\bf \large{Acknowledgements}}\\
W.-X.W. was supported by NSFC under Grant No. 11105011, CNNSF under
Grant No. 61074116 and the Fundamental Research Funds for the
Central Universities. Y.-C.L. was supported by AFOSR under Grant
No. Fa9550-10-1-0083 and by NSF under Grant No. CDI-1026710.\\

\noindent
{\bf \large{Author contributions}}\\
W.-X.W., Z.S., Y.F., Z.D. and Y.-C.L. designed research; Z.S. and W.-X.W. performed research;
Y.F. and Z.D. contributed analytic tools; Z.S., W.-X.W., Y.F., Z.D. and Y.-C.L. analyzed
data; and W.-X.W. and Y.-C.L. wrote the paper.\\

\noindent
{\bf \large{Additional information}}\\
\noindent
{\bf Supplementary Information} accompanies this paper as http://www.nature.com/
naturecommunications\\

\noindent
{\bf Competing financial interests:} The authors declare no competing financial interests.\\

\noindent
{\bf reprints and permission} information is available online at http://npg.nature.com/
reprintsandpermissions/\\

\noindent
{\bf How to cite this article:}

\begin{table}[h]\scriptsize
\centering
\caption{{\bf Performance of reconstruction and locating hidden source.}
The accuracy of network reconstruction is quantified by the success rates
SREL and SRNC as well as the trade-off measures TPR and FPR. The accuracy
in determining the values of the infection rate $\lambda$ is characterized by
the relative mean errors, the minimum
and maximum errors. The accuracy of locating hidden source (LHS)
is characterized by the trade-off measures TPR and FPR. The results of network
reconstruction and error in $\lambda$ are obtained from 30 independent
realizations. The results of locating hidden source is obtained from 10
dynamical realizations and 10 configurations of the hidden source. Other
parameters are the same as in Fig.~\ref{fig:gap}. For
data sources, reference and network models, see Supplementary Table 1
and Note~10.}
\label{tab:suc}
\begin{tabular}{*{14}{p{1.0cm}}}
\hline
\hline
\multirow{2}{*}{} &&& \multicolumn{4}{c}{Network reconstruction} & & \multicolumn{3}{c}{Error of $\lambda$} & & \multicolumn{2}{c}{LHS}\\
[0.75 ex] &&& SREL &  SRNC & TPR & FPR & & Mean & Min & Max & & TPR & FPR  \\
\hline
\multirow{10}{*}{SIS}  & \multirow{1}{*}{WS} & & 1.0 & 1.0 & 1.0 & 0.0 &  & 0.008 & 0.0 & 0.042 & & 1.0 & 0.001 \\
& \multirow{1}{*}{ER} & & 0.992 & 0.991 & 0.992 & 0.009 &  & 0.014 & 0.0 & 0.192 & & 0.977 & 0.026 \\
& \multirow{1}{*}{BA} & & 0.977 & 0.986 & 0.977 & 0.014 &  & 0.046 & 0.0 & 1.881 & & 0.99 & 0.066 \\
& \multirow{1}{*}{prison} & & 0.994 & 0.993 & 0.994 & 0.007 &  & 0.032 & 0.0 & 0.311 & & 1.0 & 0.017 \\
& \multirow{1}{*}{Santafe} & & 0.978 & 0.991 & 0.978 & 0.009 &  & 0.063 & 0.0 & 1.122 & & 1.0 & 0.041 \\
& \multirow{1}{*}{netscience} & & 0.959 & 0.992 & 0.959 & 0.008 &  & 0.172 & 0.0 & 2.878 & & 0.949 & 0.144 \\
& \multirow{1}{*}{NW} & & 1.0 & 0.999 & 1.0 & 0.001 &  & 0.015 & 0.0 & 0.15 & & 1.0 & 0.001 \\
& \multirow{1}{*}{ZK} & & 0.981 & 0.993 & 0.981 & 0.007 &  & 0.079 & 0.001 & 0.712 & & 0.97 & 0.078 \\
& \multirow{1}{*}{polbooks} & & 0.932 & 0.984 & 0.932 & 0.016 &  & 0.133 & 0.0 & 1.1875 & & 0.732 & 0.234 \\
& \multirow{1}{*}{football} & & 0.970 & 0.989 & 0.970 & 0.011 &  & 0.166 & 0.001 & 0.651 & & 0.612 & 0.021 \\
& \multirow{1}{*}{dolphin} & & 0.952 & 0.967 & 0.952 & 0.033 &  & 0.022 & 0.001 & 0.123 & & 0.925 & 0.083 \\
& \multirow{1}{*}{leadership} & & 0.975 & 0.986 & 0.975 & 0.014 &  & 0.05 & 0.0 & 0.706 & & 0.869 & 0.057 \\
 \hline
\multirow{10}{*}{CP}  & \multirow{1}{*}{WS} & & 1.0 & 1.0 & 1.0 & 0.0 &  & 0.009 & 0.0 & 0.048 & & 0.936 & 0.068 \\
& \multirow{1}{*}{ER} & & 0.999 & 1.0 & 1.0 & 0.0 &  & 0.01 & 0.0 & 0.073 & & 0.925 & 0.327 \\
& \multirow{1}{*}{BA} & & 0.997 & 1.0 & 1.0 & 0.0 &  & 0.008 & 0.0 & 0.043 & & 0.943 & 0.08 \\
& \multirow{1}{*}{prison} & & 0.995 & 0.996 & 0.996 & 0.004 &  & 0.005 & 0.0 & 0.018 & & 0.911 & 0.012 \\
& \multirow{1}{*}{Santafe} & & 0.984 & 0.996 & 0.996 & 0.004 &  & 0.006 & 0.0 & 0.036 & & 0.929 & 0.034 \\
& \multirow{1}{*}{netscience} & & 0.996 & 0.999 & 0.996 & 0.001 &  & 0.007 & 0.0 & 0.166 & & 1.0 & 0.050 \\
& \multirow{1}{*}{NW} & & 1.0 & 1.0 & 1.0 & 0.0 &  & 0.009 & 0.0 & 0.052 & & 0.98 & 0.034 \\
& \multirow{1}{*}{ZK} & & 0.992 & 0.992 & 0.992 & 0.008 &  & 0.007 & 0.001 & 0.022 & & 0.977 & 0.028 \\
& \multirow{1}{*}{polbooks} & & 0.973 & 0.995 & 0.973 & 0.005 &  & 0.008 & 0.0 & 0.042 & & 0.829 & 0.386 \\
& \multirow{1}{*}{football} & & 0.995 & 0.997 & 0.995 & 0.003 &  & 0.006 & 0.0 & 0.028 & & 0.517 & 0.015 \\
& \multirow{1}{*}{dolphin} & & 0.952 & 0.971 & 0.971 & 0.029 &  & 0.006 & 0.0 & 0.026 & & 0.786 & 0.059 \\
& \multirow{1}{*}{leadership} & & 0.996 & 0.993 & 0.993 & 0.007 &  & 0.006 & 0.0 & 0.013 & & 0.857 & 0.075 \\
\hline
\hline
\end{tabular}
\end{table}

\end{document}